\documentclass[aps, prl, twocolumn, superscriptaddress, showpacs]{revtex4}
\usepackage{amssymb}
\usepackage{amsmath}
\usepackage{bm}
\usepackage{graphicx}
\usepackage{dcolumn}
\usepackage{longtable}
\usepackage{color}

\newcolumntype{C}[1]{>{\hfil}m{#1}<{\hfil}}

\begin{document}

\title{Negative-chirality order in $S=1/2$ kagome antiferromagnet CdCu$_{3}$(OH)$_{6}$(NO$_{3}$)$_{2}\cdot $H$_{2}$O}
\author{Yoshihiko~Ihara}
\altaffiliation{yihara@phys.sci.hokudai.ac.jp}
\author{Ryoga~Hiyoshi}
\author{Masakazu~Shimohashi}
\author{Kaoru~Hayashi}
\affiliation{Department of Physics, Faculty of Science, Hokkaido University, Sapporo 060-0810, Japan}
\author{Ryutaro~Okuma}
\altaffiliation{ryutaro.okuma@physics.ox.ac.uk}
\affiliation{Clarendon Laboratory, University of Oxford, Oxford OX1 3PU, United Kingdom}
\affiliation{Institute for Solid State Physics, University of Tokyo, Kashiwa, Chiba 277-8581, Japan}
\author{G{\o}ran~Nilsen}
\affiliation{ISIS Neutron and Muon Source, Science and Technology Facilities Council, Didcot OX11 0QX, United Kingdom}
\author{Zenji~Hiroi}
\affiliation{Institute for Solid State Physics, University of Tokyo, Kashiwa, Chiba 277-8581, Japan}
\date{\today}

\begin{abstract}
The neutron diffraction and nuclear magnetic resonance (NMR) measurements have been used to microscopically analyze 
the magnetic structure in the $S = 1/2$ kagome antiferromagnet CdCu$_{3}$(OH)$_{6}$(NO$_{3}$)$_{2}\cdot $H$_{2}$O. 
Below the magnetic ordering temperature $T_N\simeq 4$ K, magnetic Bragg reflections at (110) and (100) were found in the neutron diffraction pattern, 
which suggests a $\bm{q}=0$ magnetic structure. 
Furthermore, the vector spin chirality for the $\bm{q}=0$ structure was successfully identified from the internal field direction obtianed by the $^{14}$N-NMR measurement. 
Our findings point to a chirality-ordered magnetic structure with negative vector chirality and $\langle 100 \rangle$ anisotropy. 
\end{abstract} 

\pacs{}
\maketitle

\section{introduction}
The emergence of quantum magnetism on a two-dimensional (2-D) kagome network has attracted considerable interest. 
Since a triangular unit of localized magnetic moments shares a corner with its neighbors, 
the kagome antiferromagnets can make a huge variety of energetically degenerate spin configurations. 
The kagome antiferromagnets are one of the most promising candidates for realizing the quantum spin liquid, 
owing to enhanced magnetic fluctuations on a low dimensional magnetic network. 
The spin liquid behavior has been actually observed in herbertsmithite. \cite{shores-JACS127, norman-RMP88} 
In some other materials, however, the material-specific small perturbations produce a long-range magnetic ordering.   
Experimental investigations for the ground states of the quantum kagome magnets should be conducted from microscopic viewpoint 
to make a progress in the understanding of intriguing quantum effects that appear in the kagome-based materials. 

\begin{figure}
\begin{center}
\includegraphics[width=8.5cm]{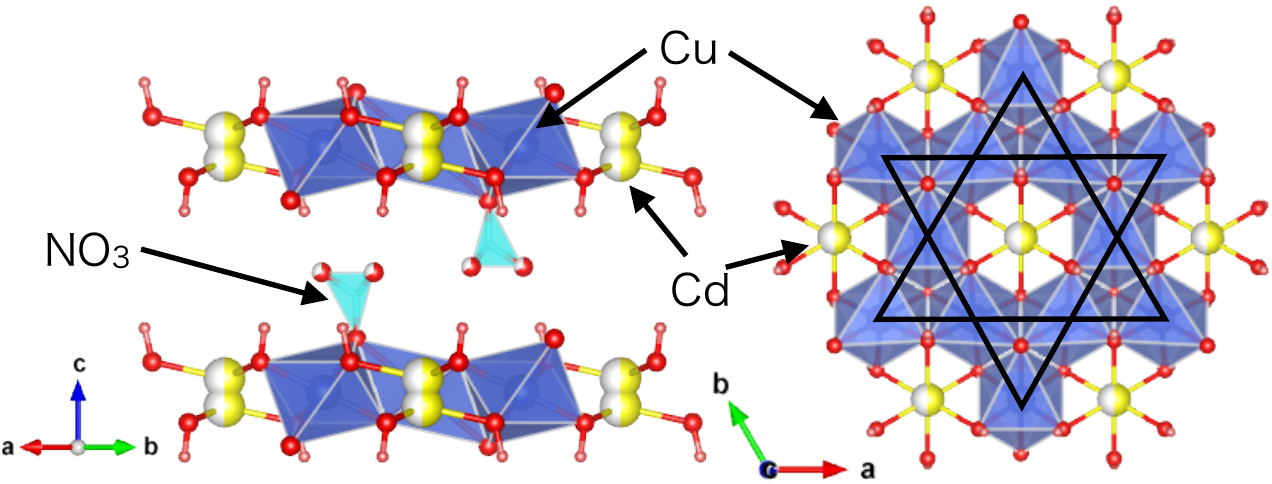}
\end{center}
\caption{
The crystal structure of CdK. Blue, yellow/white, red, cyan, and salmon spheres represent copper, cadmium, oxygen, nitrogen, and hydrogen atoms, respectively. 
The nitrate ion coordinated to the copper ion has an orientational disorder and can take three orientations randomly. 
The crystal structure was drawn by VESTA\cite{vesta}.
}
\label{crystr}
\end{figure}

In this study, we focus on a mineral CdCu$_{3}$(OH)$_{6}$(NO$_{3}$)$_{2}\cdot $H$_{2}$O (Cd-kapellasite, CdK), 
in which $S=1/2$ Cu$^{2+}$ spins form a 2-D kagome network. \cite{okuma-PRB95} 
The nonmagnetic Cd ions are located at the center of the hexagon (Fig.~\ref{crystr}) and provide the off-diagonal interaction across the hexagon $J_d$. 
With a dominant nearest neighbor and a small next nearest neighbor interactions, $J_1$ and $J_2$, 
the magnetism in CdK is described by the $J_1-J_2-J_d$ model. \cite{fak-PRL109,bernu-PRB87,iqbal-PRB92} 
Within the series of these compounds, antiferromagnetic $J_1$ was found only in CdK 
and a sister compound CaCu$_{3}$(OH)$_{6}$Cl$_{2}\cdot 0.6$H$_{2}$O (Ca-kapellasite, CaK) \cite{yoshida-JPSJ86}. 
Experimental studies on these compounds are crucial as the theoretical calculations including quantum fluctuation effects are difficult for the antiferromagnetic $J_1-J_2-J_d$ model. 
In fact, previous studies on CdK have highlighted the effects of substantial quantum fluctuations in thermal Hall effect \cite{akazawa-PRX10} 
and high-field magnetization. \cite{okuma-natcommun10}
In small magnetic fields, a magnetic anomaly with weak ferromagnetic character was observed at $T_N\simeq4$ K from the abrupt increase in susceptibility 
and broad peak in the heat capacity. \cite{okuma-PRB95} 
The temperature dependence of the thermal Hall effect observed above the magnetic ordering temperatures was interpreted 
in the framework of Schwinger-boson mean field theory\cite{akazawa-PRX10, lee-PRB91}, which suggests a finite-temperature spin liquid state with DM interactions. 
A similar spin liquid behavior was also observed in CaK at corresponding temperature range from the thermal Hall effect and $^2$D-NMR measurements. \cite{ihara-JPSJ90,doki-PRL121} 
The magnetization study in pulsed high magnetic fields revealed a series of field-induced valence-bond crystals at multiple magnetization plateaus. \cite{okuma-natcommun10} 
Although the exotic nature of quantum magnetism has been detected in CdK, 
the magnetic structure in the ground state at small magnetic fields has ramained unknown. 
Thus far, a $\bm{q}=0$ state has been suggested from the weak ferromagnetic behavior in magnetization \cite{okuma-PRB95}. 
However, various types of spin configurations can be constructed even within the $\bm{q}=0$ structures and 
theoretical study suggests that the thermal Hall conductivity would vanish at a specific spin configuration. \cite{mook-PRB99}
Microscopic measurements should be performed to determine the actual magnetic structure.

The magnetic structure possible for kagome antiferromagnet is based on the $120^{\circ}$ spin configuration,
for which the neighboring two spins always form a relative angle of $120^{\circ}$. 
A classical calculation \cite{messio-PRB83} suggests that $\bm{q}=0$ or $\sqrt{3} \times \sqrt{3}$ structures are favored as ground states at large parameter space. 
Non-coplanar cuboc structure is also propose at a restricted area around $J_d=0$. 
In the case of CaK, the negative chirality $\bm{q}=0$ structure is stabilized by the energy gain of DM interaction, \cite{ihara-PRR2, iida-PRB101, ihara-JPSJ90} 
which underlines the relevance of DM interaction in establishing a theoretical model for the kagome antiferromagnet. 

The magnetic interactions in CdK were estimated from bulk susceptibility as $ (J/{\rm K}, J_2/J, J_d/J) = (45.4, –0.1, 0.18)$. \cite{okuma-natcommun10} 
The energy scale of DM interaction was suggested as approximately 4 K, which is comparable to the long-range ordering temperature. 
According to the classical theory, these parameter sets suggest a $\bm{q}=0$ structure as the ground state. 
However, in the case of a $\bm{q}=0$ structure, two distinct spin configurations are possible with the same translation vector, 
namely the vector chirality of three spins on one triangle can be either negative (negative vector chirality: NVC) or positive (positive vector chirality: PVC). 
These various spin configurations are difficult to distinguish using neutron diffraction measurement especially when the sample mass is limited, 
and the ordered moments are reduced by the quantum fluctuations. 

The NMR measurement is an excellent tool for determining the local spin configuration. 
By examining the internal fields at the target nuclei from the NMR spectra, spin configurations around the target nuclear sites can be identified. 
The sign of vector chirality in a $\bm{q}=0$ structure is determined by the orientation of dipole fields at the three-fold axis of a triangle. 
The internal-field direction is an important indicator of the spin configuration as it eliminates the requirement for quantitative estimate of the field strength, 
which is frequently influenced by many parameters. 
While sensitive to the local spin configuration, NMR measurement is incapable of determining the global translation vector. 
Therefore, we performed the $^{14}$N-NMR measurement for CdK together with the neutron diffraction measurements. 
The complementary combination of neutron diffraction and NMR study clearly identify the magnetic structure in the ground state.

\section{Experimental}

Single crystals of CdK used for the NMR experiment were grown by the hydrothermal transport method.\cite{okuma-PRB95} 
The NMR spectra were measured in the magnetic fields $\bm{H}$ applied along $[001]$ and $[120]$ directions. 
For the sharp spectra in $\bm{H} \parallel [001]$ the full spectra were obtained by the fast Fourier transform (FFT) of the spin-echo signal at a fixed field. 
In $\bm{H} \parallel [120]$, since the spectral width becomes significantly broad by the effect of nuclear quadrupolar interaction, 
we measured the field-sweep spectra at a fixed frequency, for which the FFT intensity was recorded during the field sweep. 
We measured the temperature dependence of NMR frequency shift at high temperatures above 100 K for $H \parallel [001]$, 
but the peak positions hardly shift, suggesting that the transferred hyperfine interactions between Cu$^{2+}$ electronic spins and 
$^{14}$N nuclear spins are compensated by the direct dipole interactions, 
which were calculated to be approximately 100 mT/$\mu_B$.

\begin{figure*}[t]
\begin{center}
\includegraphics[width=14cm]{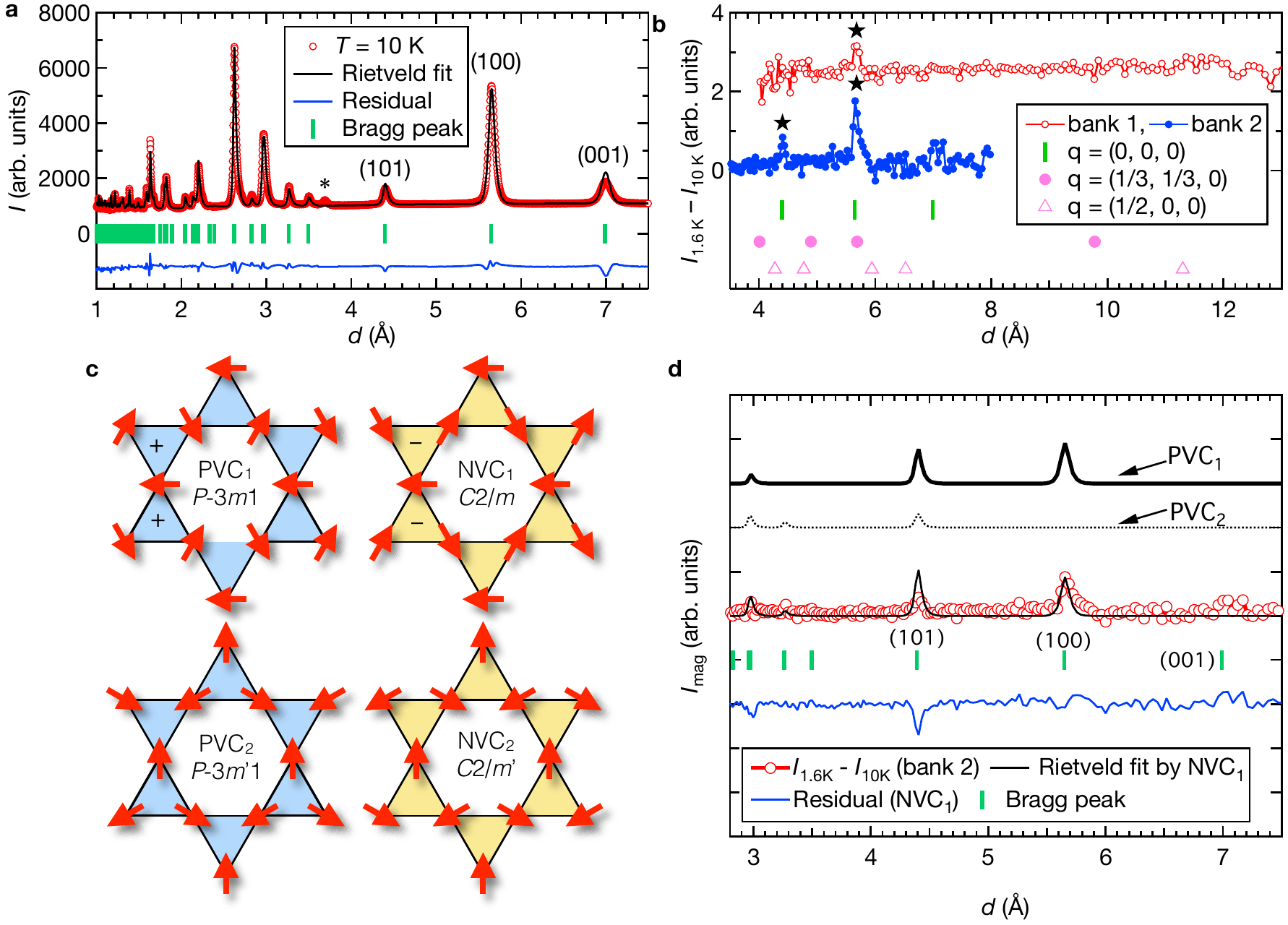}
\end{center}
\caption{
Neutron diffraction results of CdK. 
(a) The nuclear Rietveld refinement of the neutron diffraction pattern at 10 K. 
The red open circle, black line, and blue line indicate the observed, calculated and residual patterns. 
The green bars represent the positions of the nuclear Bragg peak. 
Impurity contribution shown by asterisk is excluded from the refinement. 
(b) The subtracted diffraction of the dataset at 1.6 K from the one at 10 K in bank 1 (red line with open circles) and bank 2 (blue line with filled circles) on WISH diffractometer. 
The observed magnetic Bragg peaks are shown by black stars. 
The green bar, pink filled circle, and open triangle represent the expected positions of Bragg peaks corresponding to $\bm{q} = (0, 0, 0), (1/3, 1/3, 0)$, and $(1/2, 0, 0)$. 
(c) Candidate antiferromagnetic structures with a propagation vector of $\bm{q} = (0, 0, 0)$ in the kagome lattice. 
PVC$_1$ (PVC$_2$) and NVC$_1$ (NVC$_2$) structures are the ordering of positive and negative vector chirality in the presence of $\langle 100 \rangle (\langle 120 \rangle)$ anisotropy, respectively. 
(d) Magnetic Rietveld refinement. Red open circles with thin black line, blue line, and green bar represent the observed magnetic diffraction pattern, 
the calculated pattern of NVC$_1$ structure, the residual of the fitting by NVC$_1$ structure, and the position of the Bragg peaks, respectively. 
The thick and dotted black lines indicate the calculated pattern of PVC$_1$ and PVC$_2$, respectively.
}
\label{magstr}
\end{figure*}

Powder sample of CdK for the neutron diffraction experiment was enriched with deuterium and $^{114}$Cd 
because of large incoherent scattering of hydrogen and strong neutron absorption of natural Cd. 
Details of sample preparation are presented in Supplemental Material. \cite{sup}
A magnetization measurement confirmed that the isotope-substituted powder has the same transition temperature of 4 K as in the naturally abundant CdK. 
Neutron diffraction experiments were performed on a time of flight diffractometer WISH at ISIS Neutron and Muon Source, UK. 
Data were recorded on 10 fixed angle detector banks in five pairs at 1.6 K and 10 K by using the sample of deuterated Cd in a vanadium cylinder.

\section{Results and discussion}
A powder neutron diffraction experiment was performed to determine the nuclear and magnetic structures of CdK below and above the N{\' e}el temperature of 4 K. 
The observed and calculated powder diffraction patterns at 10 K are shown in Fig.~\ref{magstr}a. 
The initial structure model was based on the parameters from the single crystal study.\cite{Oswald} 
The Rietveld refinement using Fullprof Suite \cite{Fullprof} converged with space group $P\bar{3}m1$ and successfully determined the lattice constants 
$a=b=6.5261(2)$ \AA, $c=6.9989(6)$ \AA, and all the positions of atoms including deuterium/hydrogen, of which 
schematic image is shown in Fig.~\ref{crystr}. 
The technical details of Rietveld refinement and a list of atomic positions are presented in Supplemental Material. \cite{sup}

A low-temperature pattern collected at 1.6 K showed no additional peaks, 
suggesting absence of incommensurate structures or the $\sqrt{3}\times \sqrt{3}$ cuboc/octahedral order with propagation vector of $\bm{q} = (1/3, 1/3, 0)$ or $(1/2, 0, 0)$. 
As shown in Fig.~\ref{magstr}b, subtraction of the dataset at 1.6 K from the one at 10 K revealed (100) peak at bank 2 and (101) reflection at bank 1 and bank 2 (detectors with different $d$ spacing range). 
This points to the presence of a $\bm{q} = 0$ magnetic order in CdK. 
A $\bm{q} = 0$ antiferromagnetic order in the classical Heisenberg model on the kagome lattice is characterized by either positive or negative signs of vector chirality in each triangle. 
Figure 2c illustrates four candidate magnetic structures, PVC$_1$, PVC$_2$, NVC$_1$, and NVC$_2$, 
which are represented as $\psi_1, \psi_2, 2\psi_4 + \psi_5, 2\psi_7 + \psi_9$ 
by using basis vectors of $P\bar{3}m1$ with $\bm{q} = 0$ propagation vector as described in Ref.~\cite{okuma-PRB95}. 
Comparison of the observed and calculated diffraction patterns are shown in Fig.~\ref{magstr}d. 
We note that NVC$_1$ and NVC$_2$ structures produce an identical diffraction pattern. 
PVC$_2$ structure can be excluded because of null intensity at (100) position in the simulated diffraction pattern. 
The patterns of PVC$_1$ and NVC$_1$/NVC$_2$ are consistent with the observed one and both the structures show finite intensity at (100) and (101) 
and no intensity at (001). 
Results of the refinements are summarized in Table \ref{magpar}. 
Both PVC$_1$ and NVC$_1$ structures yielded similar $R_{\rm wp}$ and $\chi^2$ values and the ordered moment is as small as 0.21(1) $\mu_B$ in NVC$_1$ structure, 
suggesting a strong fluctuation left in the ground state.

\begin{table}
\caption{
Result of magnetic Rietveld refinement.
}
\label{magpar}
\begin{tabular}{|C{25mm}|C{15mm}|C{15mm}|C{15mm}|}\hline
& $M/\mu_B$ & $R_{\rm wp}$ (\%) & $\chi^2$ \\ \hline 
PVC$_1$ & 0.19(1) & 98 & 0.725 \\ \hline
PVC$_2$ & 0.23(2) & 99 & 0.74 \\ \hline
NVC$_1$/NVC$_2$ & 0.21(1) & 98.2 & 0.727 \\ \hline
\end{tabular}
\end{table}%

To pin down the magnetic structure uniquely, NMR measurement is useful as the complementary probe. 
We use a fact that PVC and NVC yield different orientation of dipole fields at the N sites. 
For PVC configuration, the dipole fields are perpendicular to the kagome plane, 
and their magnitude is maximum for PVC$_2$ and zero for PVC$_1$. 
In the case of NVC, dipole fields with a constant magnitude appear along the kagome plane. \cite{ihara-PRR2} 
Therefore we measured the $^{14}$N-NMR spectra at low temperatures across $T_N$ in both field directions, 
namely $[120]$ and [001] directions.

\begin{figure}
\begin{center}
\includegraphics[width=7.5cm]{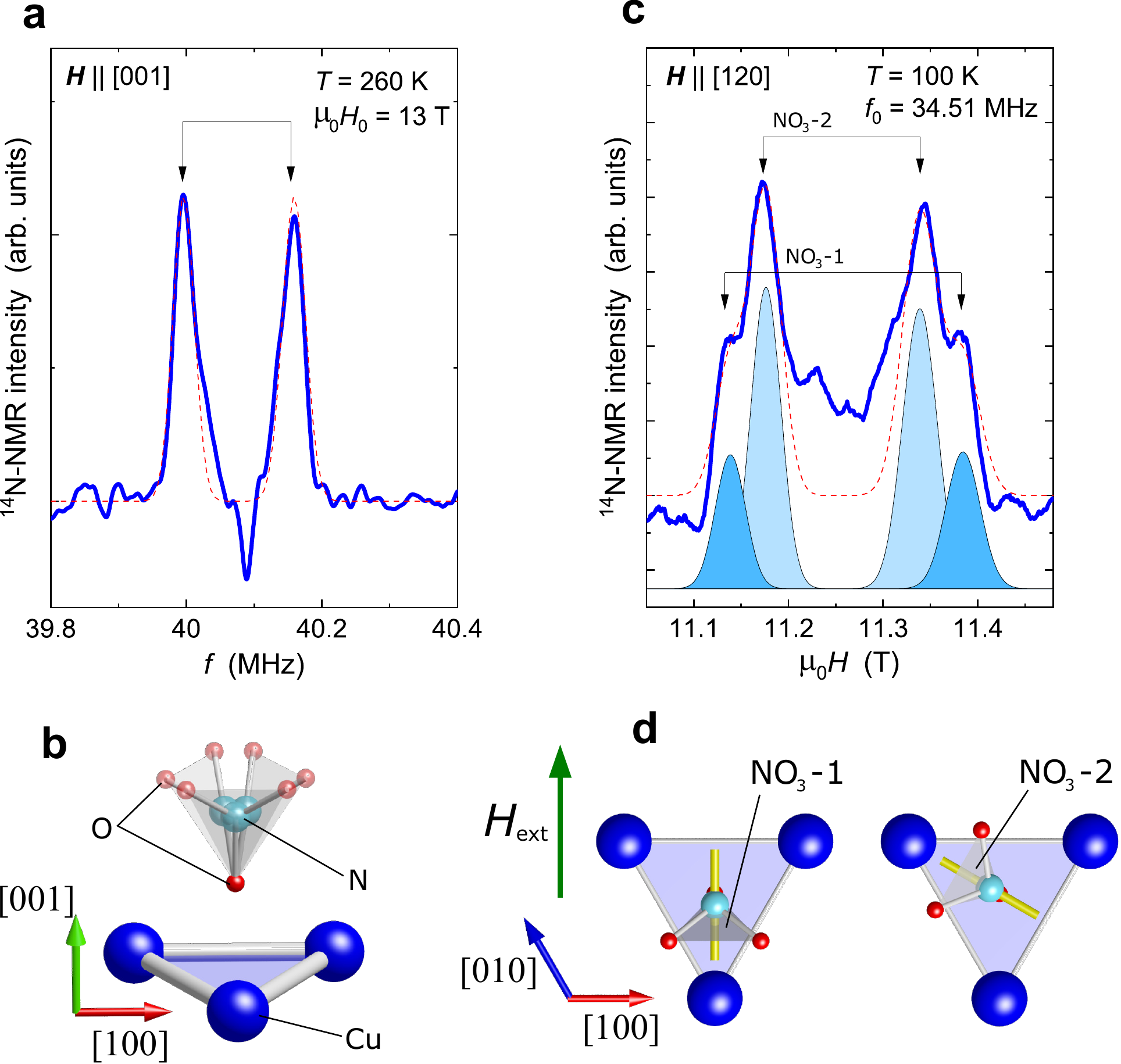}
\end{center}
\caption{
$^{14}$N-NMR spectra for single crystalline CdK at high temperatures measured in (a) $\bm{H} \parallel [001]$ and (c) $\bm{H}\parallel [120]$. 
Two-peaks structure typical for $I=1$ nuclear spin is observed in (a), 
while four peaks were observed in (c) because of the positional disorder of NO$_3$ ions. 
The shadowed four peaks are the result of spectrum simulation. (see text) 
The sum of these four peaks is represented by red dotted line. 
(b) Positional disorder of NO$_3$ ions located on a Cu triangle. Three possible NO$_3$ sites are randomly occupied. 
(d) In-plane structure of NO$_3$ ions. 
The orientation of the main principal axis of EFG is indicated by a bar on the N sites. 
When the external field $H_{\rm ext}$ is applied along $[120]$ direction, 
N sites at NO$_3$-1 and NO$_3$-2 show different quadrupolar splitting. 
}
\label{highTsp}
\end{figure}

The 2-D kagome layers are separated by NO$_{3}$ ions and the N sites are located near the center of Cu triangle as shown by the enlarged picture in Fig.~\ref{highTsp}b. 
The positional disorder of N sites creates additional feature for the NMR spectra through the electric quadrupolar interaction in fields parallel to the kagome plane. 
In contrast, when the field is perpendicular to the kagome plane, positional disorder of N sites does not split the NMR spectra 
because these N positions are reproduced by the three-fold symmetry around the [001] direction. 
Figure \ref{highTsp}a shows the $^{14}$N-NMR spectrum at room temperature in the fields parallel to the [001] direction. 
As the nuclear spin of $^{14}$N is $I =1$, two peaks from $m=1\leftrightarrow0$, and $m=0\leftrightarrow-1$ transitions were observed. 
From the separation of these two peaks, the NQR frequency along [001] direction is estimated as 164 kHz. 
The lattice symmetry imposes that one of the principal axes of electric field gradient (EFG) is perpendicular to the mirror plane that includes the target N sites 
and other two axes are in the mirror plane. 
As the three covalent N-O bonds dominantly contribute to the EFG, the main principal axis that has the largest EFG should be perpendicular to the O triangle of NO$_3$ unit 
as shown in Fig.~\ref{highTsp}d. 
The randomly occupied three N sites generate three different orientations of EFG principal axes in the $ab$ plane. 
Thus, for the inplane fields, the orientational disorder of NO$_3$ units introduces an additional feature in the $^{14}$N-NMR spectrum. 
Figure~\ref{highTsp}c shows the $^{14}$N-NMR spectrum for $\bm{H} \parallel [120]$, in which we found two sets of two peaks. 
The peaks with larger splitting originate from the N sites at EFG principal axis parallel to the external fields (NO$_{3}$-1 in Fig.~\ref{highTsp}d), 
while those with narrower splitting are from NO$_3$ at other positions (NO$_{3}$-2). 
We estimated the NQR frequency along main $[120]$ direction from the peak separation for NO$_3$-1 site as 790 kHz. 
The splitting for NO$_3$-2 sites is then calculated as 0.145 T, \cite{sup}
which is in perfect agreement with the experimentally obtained peak separation of 0.16 T. 
The peak intensity of NO$_3$-2 sites are larger than that for NO$_3$-1 sites because the number of sites is twice more. 
We simulated the $^{14}$N-NMR spectrum for $\bm{H} \parallel [120]$ using the parameters obtained above. 
The result is shown in Fig.~\ref{highTsp}c as the shadowed peaks and dotted line. 
The discrepancy around the center of four peaks is coming from the imperfect orientational order of NO$_3$ units. 

\begin{figure}
\begin{center}
\includegraphics[width=7cm]{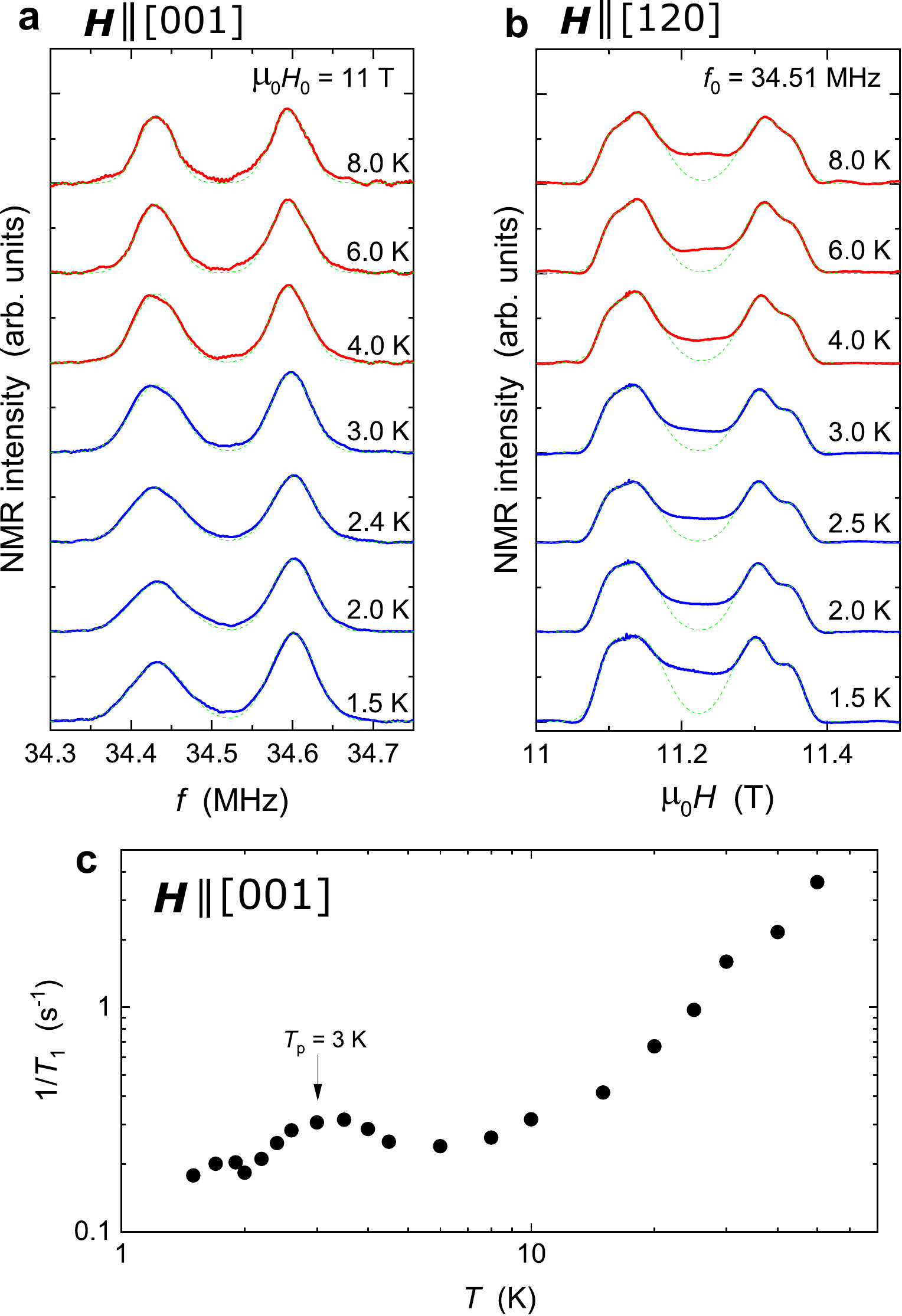}
\end{center}
\caption{
Temperature dependence of $^{14}$N-NMR spectra in (a) $\bm{H} \parallel [001]$ and (b) $\bm{H}\parallel [120]$. 
The dotted green lines are the results of multi-peak Gaussians fit. 
(c) Temperature dependence of $1/T_{1}$ in $\mu_0H \simeq 11$ T. 
A peak in $1/T_{1}$ at $T_N=3$ K confirms a magnetic transition at $\sim 11$ T, 
even though no apparent change was observed in the spectral shape. 
}
\label{lowTsp}
\end{figure}

\begin{figure*}
\begin{center}
\includegraphics[width=16cm]{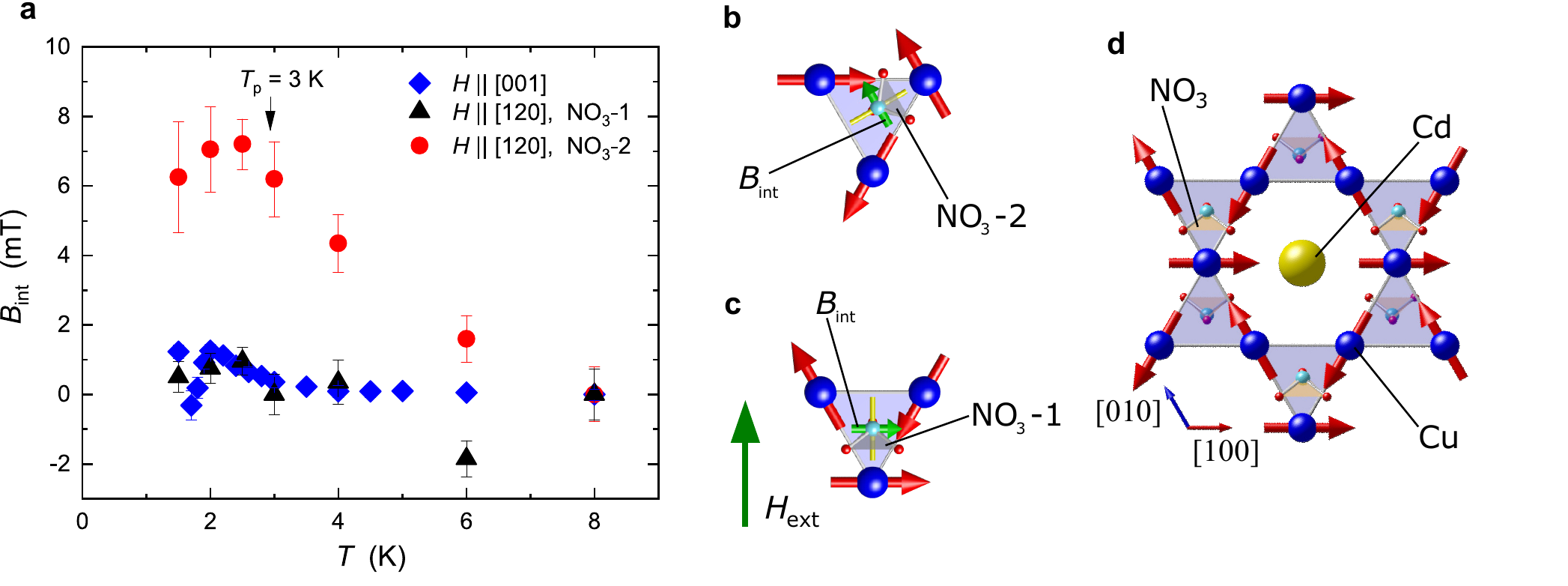}
\end{center}
\caption{
(a) Internal fields obtained from NMR shift. 
Negligibly small internal fields were observed at low temperatures in $\bm{H} \parallel [001]$ and $\bm{H}\parallel [120]$ for NO$_3$-1 sites. 
(b), (c) Negative-chirality spin configuration for Cu spins and dipole fields at the N site. 
The direction of internal fields is indicated by an arrow on the N site. 
$B_{\rm int}$ at NO$_3$-1 site is perpendicular to $H_{\rm ext}$, while the angle between $B_{\rm int}$ and $H_{\rm ext}$ is $30^{\circ}$ at NO$_3$-2 site. 
(d) Orientation-ordered crystal structures with anisotropy axis along [100] directions. 
NO$_3$ unis are located above and below the kagome plane alternatively at each triangle, 
and are reproduced by the inversion symmetry within a unit cell. 
Magnetic structures for both orientations are represented by red arrows on the Cu sites. 
Among three spins on one triangle, one spin is always parallel to the anisotropy axis. }
\label{peakshift}
\end{figure*}

Knowing the above peak assignments, we measured the temperature variation of $^{14}$N-NMR spectra passing through $T_N$.
Figures \ref{lowTsp}a, b show the NMR spectra for both field directions below 8 K. 
Although only a small modification was observed on the spectral shape around $T_N$,
we confirmed the magnetic transition temperature at approximately 11 T by the $1/T_1$ measurement in $\bm{H} \parallel [001]$. 
As shown in Fig.~\ref{lowTsp}c, a peak in $1/T_1$ was observed at $T_N=3$ K, which is associated with the critical magnetic fluctuations near the magnetic phase transition. 
$T_N$ in 11 T is slightly lower than that at zero fields because the antiferromagnetically ordered state is suppressed in magnetic fields. 
Across the magnetic phase transition at $T_N$, no change in spectral shape was observed for $\bm{H} \parallel [001]$. 
The spectra in $\bm{H} \parallel [120]$ show a shift in the peak position only for the NO$_3$-2 sites, which results in the asymmetric spectral shape at the lowest temperature.

The temperature dependence of $^{14}$N-NMR shift is determined by fitting the NMR spectra with multi-peak Gaussians 
and the results for three independent $^{14}$N sites are shown in Fig.~\ref{peakshift}a. 
The internal fields along the external field directions are estimated by taking the average of peak positions for the quadrupolar-split two peaks. 
We defined $\Delta B_{\rm int}(T) = B_{\rm int}(T)-B_{\rm int}(8$ K). 
A significant temperature dependence is observed only for NO$_3$-2 sites in $\bm{H} \parallel [120]$, 
which leads us to conclude the negative chirality $\bm{q}=0$ structure as we will discuss next. 

In the $\bm{q}=0$ state with 120$^{\circ}$ spin configuration, the hyperfine fields from ordered moments are canceled at N sites. 
Therefore, we assume the direct dipole fields as the origin of NMR shift in the ordered state. 
The PVC spin configuration is excluded from the absence of internal fields along the $c$ direction as we explained above. 
In the present study we found that even in $\bm{H} \parallel [120]$ the internal fields are absent at NO$_3$-1 sites. 
This site dependence is explained by the spin-locking in the $\bm{q}=0$ structure. 
The direction of the internal fields depends on the global spin orientation in the NVC structure. 
When the spin direction with respect to the EFG principal axis is locked, the internal field at N site always has a fixed angle against the EFG axis 
as shown in Figs.~\ref{peakshift}b, c. 
In the case of NVC$_1$ configuration shown in Fig.~\ref{peakshift}, 
the internal fields are parallel to the kagome plane and perpendicular to the EFG axis, 
thus have no effect on the NMR shift when the external fields are parallel to the main EFG axis, which is the case for NO$_3$-1 sites. 
A finite NMR shift by the internal fields can be observed only at NO$_3$-2 sites, for which the angle between the external and internal fields is 30$^{\circ}$. 
On the other hand, when all spins are rotated by 90$^{\circ}$ (NVC$_2$), the NMR shift should be observed for the sites with fields parallel to the EFG axis, 
that is, NO$_3$-1 sites with the largest quadrupolar splitting. 
As no internal fields were observed at NO$_3$-1 sites, 
we strongly suggest a spin-locked NVC$_1$ structure as the ground state.

Now we estimate the size of the ordered moments from the internal fields in the ordered state $\Delta B_{\rm int}$. 
As the internal fields projected to the external field direction is $\sim7$ mT at $T<T_N$, the magnitude of dipole field is estimated as $7/\cos(30^{\circ})=8$ mT. 
When full moments of $1 \mu_B$ form the NVC spin configuration in CdK, the dipole fields at N sites are computed as 40 mT. 
By comparing this estimate and experimentally obtained $\Delta B_{\rm int}$, the size of the ordered moment can be estimated as $0.2 \mu_B$. 
This is consistent with the value obtained from the neutron diffraction measurement. 

The NVC state was reported in CaK from $^{35}$Cl and $^{2}$D NMR studies. \cite{ihara-PRR2, ihara-JPSJ90} 
In CaK, however, spin directions are not locked to the crystal axes. 
The resulting NMR spectra in the ordered state show a broadening rather than a uniform shift. 
By contrast, negligibly small spectral broadening in CdK clearly suggests the spin-locked state. 
As the spin direction is always fixed with respect to the EFG principal axis, we suggest that the orientation of  NO$_3$ is long-range ordered, which is responsible for the spin locking. 
Small modification of Cu-O-Cu bonding through O of the NO$_3$ units would introduces bond-dependent exchange matrices, 
which then produce the magnetic anisotropy with respect to $\langle 100\rangle$ direction. 

The simplest ordering of the orientation of NO$_3$ compatible with $\bm{q} = 0$ structure and the spin locking is plotted in Fig.~\ref{peakshift}d, which belongs to $C2/m$ space group as found by the symmetry analysis using ISODISTORT\cite{isodistort1, isodistort2, sup}. 
This monoclinic structure is supposed to coexist with the other two domains related by three-fold rotation with the equal probability. 
While the limited crystallinity of the powder sample prevents us from elucidating the long-range order of NO$_3$ by neutron scattering \cite{sup}, 
the spin orientations of the NVC$_1$ structure are uniquely determined within a specific domain by the orientation of NO$_3$ units.
Therefore, the uniform internal field appears at the N sites throughout a single domain and its direction is parallel to the anisotropy axis, 
which is consistent with the uniform shift of the entire NMR spectra only for NO$_3$-2 sites.

To summarize, we have performed the neutron diffraction and $^{14}$N-NMR measurements of 2-D kagome antiferromagnet CdK and 
identified the $\bm{q}=0$ state with the chirality-ordered NVC$_1$ structure. 
The $\bm{q}=0$ state is consistent with the classical calculation with $J_1-J_2-J_d$ model, 
but in CdK DM interaction and modified exchange interaction due to orientational order of NO$_3$ units are also important to stabilize the spin-locked NVC$_1$ structure. 
The ordered moment is significantly reduced even at the lowest temperature because of strong quantum fluctuations. 
Since the magnetic ground state has been unambiguously determined, 
fascinating phenomena in CdK would be understood in terms of quantum fluctuations activated on the NVC$_1$ background. 
In fact, suppression of thermal Hall conductivity toward the ordered state \cite{akazawa-PRX10} is consistently explained by the NVC$_1$ structure, 
because the magnetic symmetry prohibits the thermal Hall effects in the case of NVC$_1$. \cite{mook-PRB99}

\begin{acknowledgements}
We would like to acknowledge R. Kumar, J. Ohara, and H. K. Yoshida for fruitful discussions and  F. Orlandi for helping the neutron diffraction experimentsy.
R.O. was supported by the Materials Education Program for the Future Leaders in Research, Industry, and Technology (MERIT) under MEXT. 
The study was partially supported by KAKENHI (Grant Nos. 15K17701, 18H01163, 19H01832, 21H01035), 
the Core-to-Core Program for Advanced Research Networks under JSPS.
\end{acknowledgements}

\end{document}